\documentclass[10pt]{iopart}

\usepackage{iopams}  
\usepackage{graphicx}
\usepackage{xcolor}

\begin{document}

\title{Nanophotonic boost of intermolecular energy transfer}

\author{P. M. de Roque}
\address{ICFO-Institut de Ciencies Fotoniques, Mediterranean
Technology Park, 08860 Castelldefels (Barcelona), Spain}

\author{N. F. van Hulst}
\address{ICFO-Institut de Ciencies Fotoniques, Mediterranean
Technology Park, 08860 Castelldefels (Barcelona),
Spain}
\address{ICREA-Instituci\'{o} Catalana de Recerca i
Estudis Avan\c{c}ats, 08010 Barcelona, Spain}

\author{R. Sapienza}
\address{Department of Physics, King's College London, Strand, London WC2R 2LS, United Kingdom}
\ead{riccardo.sapienza@kcl.ac.uk}
\vspace{10pt}
\begin{indented}
\item[]April 2015
\end{indented}

\begin{abstract}
We propose a scheme for efficient long-range energy transfer between two distant light emitters separated by more than one wavelength of light, i.e. much beyond the classical F\"{o}rster radius. A hybrid nanoantenna-waveguide system mediates the transmission of energy, showing enhancements up to $10^8$ as compared to vacuum. 
Our model shows how energy transfer in nanostructured media can be boosted, beyond the simple donor Purcell enhancement, and in particular for large donor-acceptor separations. The scheme we propose connects realistic emitters and could lead to practical on-chip implementations.
\end{abstract}

\maketitle
%
%
%
%
%

\section{Introduction}

Energy transfer between an excited two level quantum emitter and an equivalent one in its ground state is a key process in many physical systems. It is found in naturally occurring living organisms which perform photosynthesis \cite{Photosynthesis_book}, and is widely exploited in man-made devices for lighting \cite{Lighting,Lighting_Nanoletters}.  Due to its sharp inverse sixth-power distance dependence as predicted by F\"{o}rster theory \cite{Foster}, it is also often used for bioimaging as an optical ruler to assess the co-presence of biomolecular probes \cite{bioFRET,Roy}.  Despite the wide range of applications profiting from it, energy transfer between two emitters is often an inefficient process, as it relies on the spatial and spectral overlap of their radiation patterns, as well as their mutual orientations. Efficient long range energy transfer could revolutionise the way we harvest solar energy \cite{Scholes_LH,NatPhotFRETSolarcells}, open new avenues for superresolution techniques \cite{NatMethFret}, and provide new sensing\cite{NatMat_FRET_Sensor} or lightning platforms \cite{Nature_FRETLight}. Furthermore, it could allow efficient wireless energy transfer \cite{SoljacicScience,Soljacic_AnnPhys} and the implementation of quantum information protocols in realistic platforms \cite{PRL_BarencoDeutsch}. 

Engineered nanophotonic materials largely modify the local density of optical modes (LDOS) available to an emitter and hence its spontaneous emission rate \cite{Purcell}. The LDOS varies with the emitter position in an inhomogeneous environment, and can be obtained from the dyadic Green's function, via its imaginary part $\mathrm{Im} [\mathbf{G}(\mathbf{r},\mathbf{r})]$, at the emitter location $\mathbf{r}$. Energy transfer, instead, is related to $\mathbf{G}(\mathbf{r}_D,\mathbf{r}_A)$, function of both the donor ($\mathbf{r}_D$) and acceptor position ($\mathbf{r}_A$).
In homogenous media a unified quantum electrodynamical theory for radiationless (mediated by virtual photons) and radiative (mediated by a real photon) energy transfer has been developed \cite{Andrews_1989}, based on dipole-dipole interactions. A similar treatment cannot be easily applied to nanostructured media \cite{Andrews_2000} where analytical studies exist for simple geometries such as multilayers and microspheres \cite{theoryETLDOS}.
Beyond F\"{o}rster theory, an equivalent of the Purcell effect for energy transfer in nanostructured media, i.e. a dependence of the energy transfer rate on the LDOS, has been expected to follow a quadratic dependence \cite{theoryETLDOS2} or even to be independent on the LDOS \cite{theoryETLDOS3,theoryETLDOS4}. This has fuelled an ongoing debate on the role of LDOS on energy transfer: energy transfer experiments in dielectrics and moderately small emission rate enhancements have  shown energy transfer rates which do not depend on the LDOS \cite{Blum,Rabouw,Konrad}. On the contrary, for the case of plasmonic nano-apertures or films, with larger donor-acceptor distances and stronger LDOS enhancements, energy-transfer efficiencies which depend on the LDOS of the photonic environment have been reported \cite{J.Wenger,Barnes,Torres}. The role of large LDOS gradients and of nanostructured media on energy transfer is still unclear and a flexible and practical theoretical method embracing them has so far been lacking.
\begin{figure}
\begin{center}
\includegraphics[width=3in]{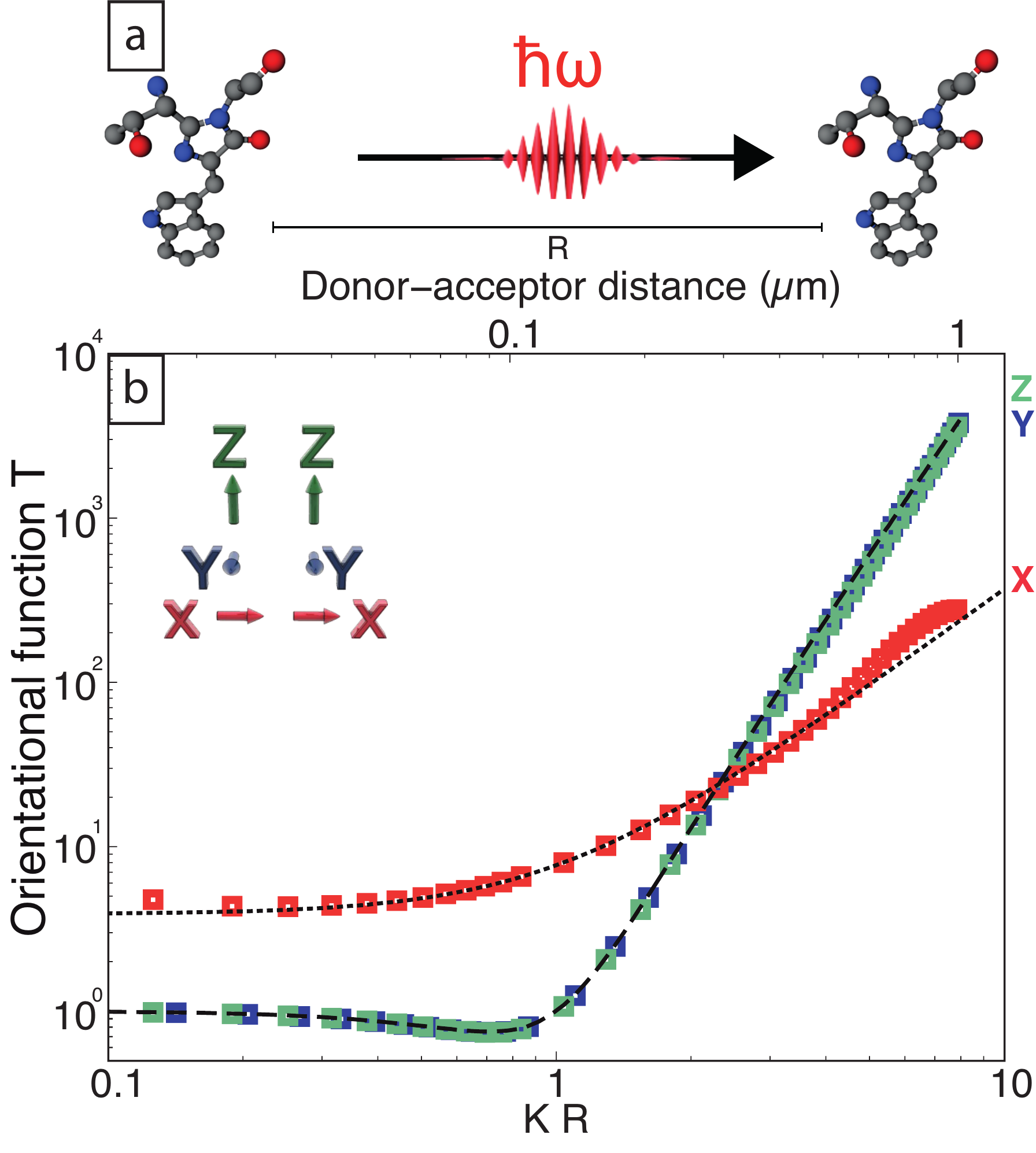} 
\caption{\label{Fig.1} (Color online) (a) Schematic illustration of energy transfer between two equivalent molecules. (b) Agreement between our numerical (symbols) and  F\"{o}rster theory (dashed line for parallel dipoles and dotted line for collinear dipoles) calculation of energy transfer for  dipole pairs  in vacuum separated by a distance $R$ ($\lambda$ = 800 nm). The orientational factor $T(\omega)$ between a donor and an acceptor oriented along each of the three cartesian axis, labelled as in the inset, is shown. } 
\end{center}
\end{figure}

Here we propose a novel scheme for long-range energy transfer based on near-field (via plasmonic antennas) and far-field (via dielectric waveguides) engineering, which can boost  the energy transfer between two emitters at a distance of several wavelengths, far beyond the few-nm F\"{o}rster radius, by 8 orders of magnitude. Using finite-difference time-domain (FDTD) modelling, a 3D vectorial numerical method including retardation effects and the full metal losses \cite{Taflove}, we compute energy transfer in the presence of arbitrary inhomogeneous environments by calculating both the Green's function and induced polarisability of the acceptor. 

\section{Energy transfer in vacuum: validation of the numerical method.}

\begin{figure}
\begin{center}
\includegraphics[width=6in]{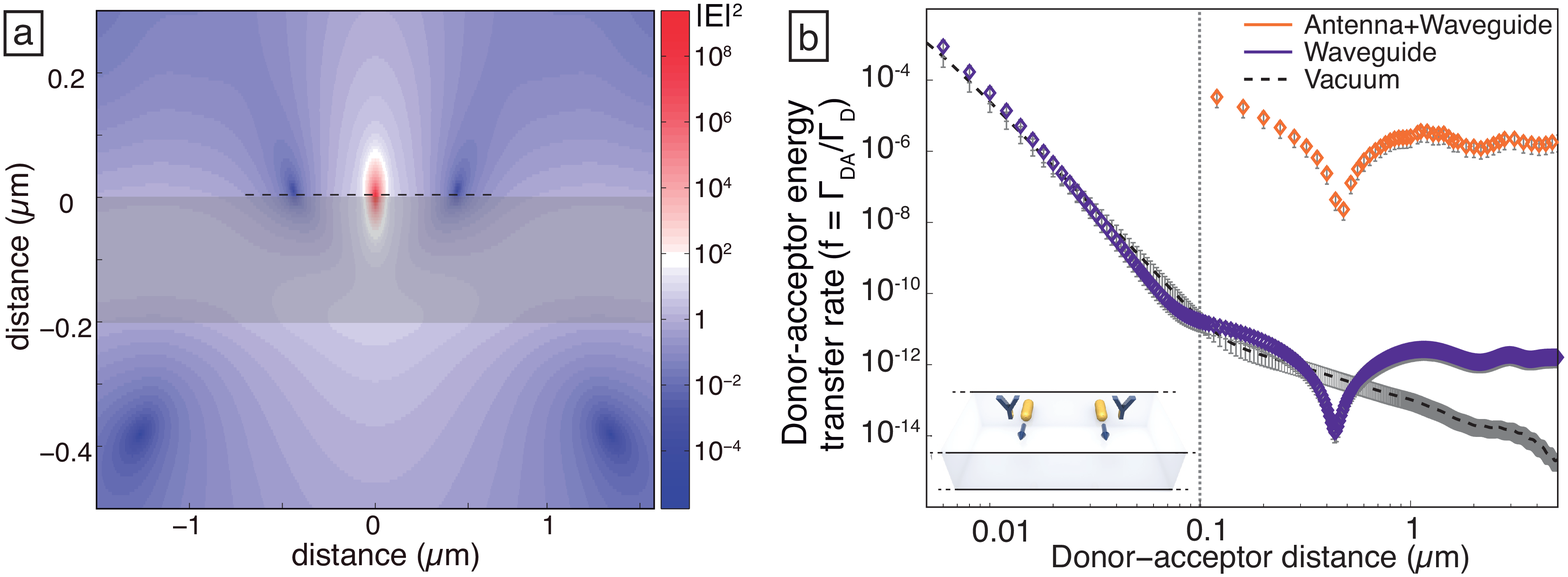} 
\caption{\label{Fig.2}  (Color online) Inhomogeneous environments modify energy transfer rate between a donor and an acceptor. (a) Time-integrated electric field intensity (arbitrary units) created by a dipole placed 10 nm on top of a 200 nm-thick dielectric membrane (n = 2). The dashed line represents the position at which the quantities in (b) are evaluated. (b) Energy transfer rate per donor emission event ($f = \Gamma_{DA}/\Gamma_D$) for two parallel dipoles placed in (1) vacuum (dashed line), (2) over a waveguide ($\textcolor[rgb]{0.5,0.0,0.99}{\boldsymbol{\Diamond}}$) and (3) next to metallic antennas over a waveguide ($\textcolor[rgb]{0.99,0.5,0}{\boldsymbol{\Diamond}}$). }
\end{center}
\end{figure}

The energy transfer rate between a donor ($\boldsymbol{\mu}_D$) and an acceptor ($\boldsymbol{\mu}_A$), as sketched in Fig. 1a,   can be expressed, when normalised by the donor decay rate,  as 
\begin{equation}
f(\omega) = \frac{\Gamma_{DA}}{\Gamma_D} = \frac{\epsilon_0 \mathrm{Im}[\boldsymbol{\alpha}_A(\omega)] |\boldsymbol{\mu}_D \cdot \mathbf{G}(\mathbf{r}_D,\mathbf{r}_A,\omega) \boldsymbol{\mu}_A|^2}{\mathrm{Im}[\boldsymbol{\mu}_D \cdot \mathbf{G}_D(\mathbf{r}_D,\mathbf{r}_D,\omega) \boldsymbol{\mu}_D]},
\end{equation}
where we denote by $(\mathbf{r}_A,\boldsymbol{\mu}_A)$ the position and the direction of the transition dipole of the acceptor (and equivalently for the donor),  $\epsilon_0$ is the vacuum permittivity, $\mathbf{G}(\mathbf{r}_D,\mathbf{r}_A,\omega)$ is the dyadic Green's function describing the photonic environment, and $\boldsymbol{\alpha}_A(\omega)$ is the polarisability of the acceptor at a frequency $\omega$  \cite{Novotny_Hecht}. 

While the energy emitted by a dipole depends on the dipole's own scattered-field emitted at a former time, energy transferred from the donor to the acceptor instead depends on both the donor field $\mathbf{E}(\mathbf{r}_A)$ and the acceptor dipole moment that this field induces, which is $\mathbf{p}_A = \boldsymbol{\alpha}_A(\omega)\mathbf{E}(\mathbf{r}_A)$; in our notation $\boldsymbol{\mu}_A = \mathbf{p}_A/|\mathbf{p}_A|$.
The Foster rate $\Gamma_{DA}$ in the homogenous medium with real and positive permittivity of the medium ($\epsilon_m$) scales as $\epsilon_m^{2}$ \cite{Foster}, while the function $f(\omega)$ in Eq. 1 scales as  $f(\omega) \propto \epsilon_m^{-5/2}$, coming from the dependence of the Green function $\mathbf{G}(\mathbf{r}_D,\mathbf{r}_A) \propto \epsilon_m^{-1}$ in the near-field and of the Purcell factor $\mathrm{Im}[\mathbf{G}(\mathbf{r}_D,\mathbf{r}_D)] \propto \sqrt{\epsilon_m}$, as the polarisability $\alpha(\omega)$ is independent on $\epsilon_m$. 
In homogenous optical media \cite{Albaladejo,deVries}, simple inhomogeneous environments like multilayers \cite{ET_multilayer} or spherical particles \cite{Carminati_magneto}, radiative corrections to the electrostatic polarisability have been introduced analytically and the energy transfer can be calculated. Solutions to Eq.(1) for lossy and more complex structures require more care and are often tackled by defining quasi-normal modes \cite{Kristensen}, or treated using the quasi-static approximation \cite{Carminati_CDOS}. Instead, our technique to calculate energy transfer based on FDTD (see Supplementary Material), can be applied vectorially to arbitrarily complex inhomogeneous media taking into account any retardation effects. Eq. 1 shows that the energy transfer can be boosted by increasing either the polarisability at the acceptor position $\boldsymbol{\alpha}_A(\omega)$ via near-field enhancement, or by engineering how the field emitted by the donor reaches the acceptor via $\mathbf{G}(\mathbf{r}_D,\mathbf{r}_A,\omega)$.

In vacuum (or homogenous media), F\"{o}rster theory analytically computes the energy transfer by defining the \emph{orientational} T factor as $T (\omega) = R^6 (\Gamma_{DA}/\Gamma_D)$ which is expressed as: $T (\omega) = 16 \pi^2 k^4 R^6 |\boldsymbol{\mu}_D \cdot \mathbf{G}(\mathbf{r}_D,\mathbf{r}_A,\omega) \boldsymbol{\mu}_A|^2$, where $k$ is the wavenumber at the frequency $\omega$ and $R = |\mathbf{r}_D - \mathbf{r}_A|$\cite{Novotny_Hecht}. In Fig. 1b we compare the analytical values with our numerical results for dipole pairs oriented along the three cartesian axis.  Evaluation of $T(\omega)$ for values of $k\cdot R \leq 1$ shows an almost flat line, hence a $R^{-6}$ dependency of the energy transfer. Additionally, energy transfer depends on the relative orientation of the dipoles. Non-collinear but parallel dipoles (y-y and z-z) show a transfer rate that drops at intermediate distances decaying as $R^{-4}$, characteristic of an inductive intermediate field. Further apart, the energy transfer trend develops into the usual $R^{-2}$ dependency when mediated by transversal photons. Instead, for collinear dipoles (x-x) the energy transfer rate decays monotonically as $R^{-4}$. Analytical and numerical values agree with an accuracy which can be increased arbitrarily at the cost of computational time and memory; here we used  a grid discretisation of size $\lambda/100$ (see Supplementary Information Fig. S1). 

\section{Energy transfer in nanostructured media}

%
%
As we show here, a hybrid waveguide-antenna system can modify long-range energy transfer from one dipolar emitter to another. The waveguide  is a free standing silicon nitride membrane, 200 nm thick and 400 nm wide, with refractive index equal to 2 supporting only one single TE mode  \cite{Yariv}. Dipoles representing donor and acceptor are placed 10 nm above the waveguide at its central position. The electric field intensity generated by a y-oriented dipole couples well to the TE waveguide mode, which is also y-polarised (Fig. 2a). Fig. 2b presents direct comparison of the energy transfer between two y-oriented dipoles for three cases: (1) in vacuum, (2) over the dielectric waveguide, and (3) over the dielectric waveguide and coupled to metallic antennas. Fig 2b plots the frequently used parameter $f = \Gamma_{DA}/\Gamma_D$, which compares the energy transfer rate to other decay mechanisms. Few regions can be identified depending on the emitter-acceptor distance. Energy transfer for distances shorter than 100 nm decays as $R^{-4}$, characteristic of coupling mediated by the intermediate-field. For larger distances, the differences between vacuum and the waveguide become more evident. At a distance of roughly 1 $\mu$m from the donor position, a lossless waveguide mode has developed, which efficiently delivers the light to the acceptor with minimal propagation loss. As a consequence, the transfer rate stops decaying for increasing donor-acceptor distances and instead saturates at a value of $\sim 2.5\cdot10^{-12}$ transferred photons per donor-emitted photon. This corresponds to an energy transfer enhancement of around 25 times when compared to vacuum at a distance of 1 $\mu$m; this is an arbitrary distance we have chosen to provide a realistic value as a guide to the experiments. The dip in Fig. 2b, at roughly 400 nm distance, is due to the way the radiation of the dipole is fed into the waveguide modes, as the dipole emission peaks at the critical angle of the air-silicon nitride interface \cite{Novotny_Hecht}. We note here that for a dipole oriented along the y-direction, only $\sim 20\%$ of the total radiation is coupled into the waveguide mode (at a wavelength of 800 nm), and only half of this ($\sim 10\%$) is coupled into the direction of the acceptor, similar to previous predictions \cite{Hwang}. Although the dielectric waveguide shows an increased long range energy transfer as compared to the vacuum case, the total rate is still limited to $\le 10^{-11}$ transfer events per donor emission event (at distances  $\geq 1\mu$m) which is far too low for any potential application with real single photon emitters. 

Efficient long-range energy transfer pathways must connect donor and acceptor near-fields. 
A waveguide mode is an excellent solution for large distances, but it is not effective in the near-field  where it overlaps weakly with the dipolar field distribution. Resonant plasmonic antennas can boost the emitter dipole moment  and enhance both emission and absorption of light \cite{antennas}; moreover, they can redirect the radiation of the emitter into specific optical modes \cite{CurtoScience}, leading to controlled scattering of the waveguide mode to match the dipole field. Our strategy is to place a plasmonic nano-antenna on the surface of the waveguide near the emitter. This hybrid system can out-perform the pure waveguide by (i) increasing near-field coupling of the donor with the waveguide, (ii) enhancing the donor decay rate by Purcell enhancement and (iii) boosting acceptor polarisability by local field engineering.  We consider a configuration in which the donor and acceptors are placed 10 nm above the structure surface and 10 nm away from the end of a 100 nm long 40 nm wide gold antenna (inset to Fig. 3b). In this new configuration the absolute energy transfer rate reaches a value of $\sim 2.5\cdot10^{-6}$ transfer events per donor emission event for distances larger than 1 $\mu$m (Fig. 2b). This is an increase of 6 orders of magnitude when compared to the waveguide alone and over 8 orders of magnitude when compared to vacuum. 

In a realistic experiment, for example with two stable molecules or quantum dots which can emit up to 1-10 Mphoton s$^{-1}$, the rate of transferred photons in the presence of our waveguide-antenna structure would then be $\sim$100-1000 photon s$^{-1}$ (assuming 100\% donor-acceptor spectral overlap). This transfer rate could potentially be measured with a state-of-the-art optical setup. 

\section{Origin of the long-range energy transfer enhancement}
We additionally explore the distance dependence of energy transfer, and show that its increase in the presence of our nanophotonic environments is due to both an increase in the donor emission rate as well as an increased number of optical modes connecting efficiently donor and acceptor.

\begin{figure}
\begin{center}
\includegraphics[width=6in]{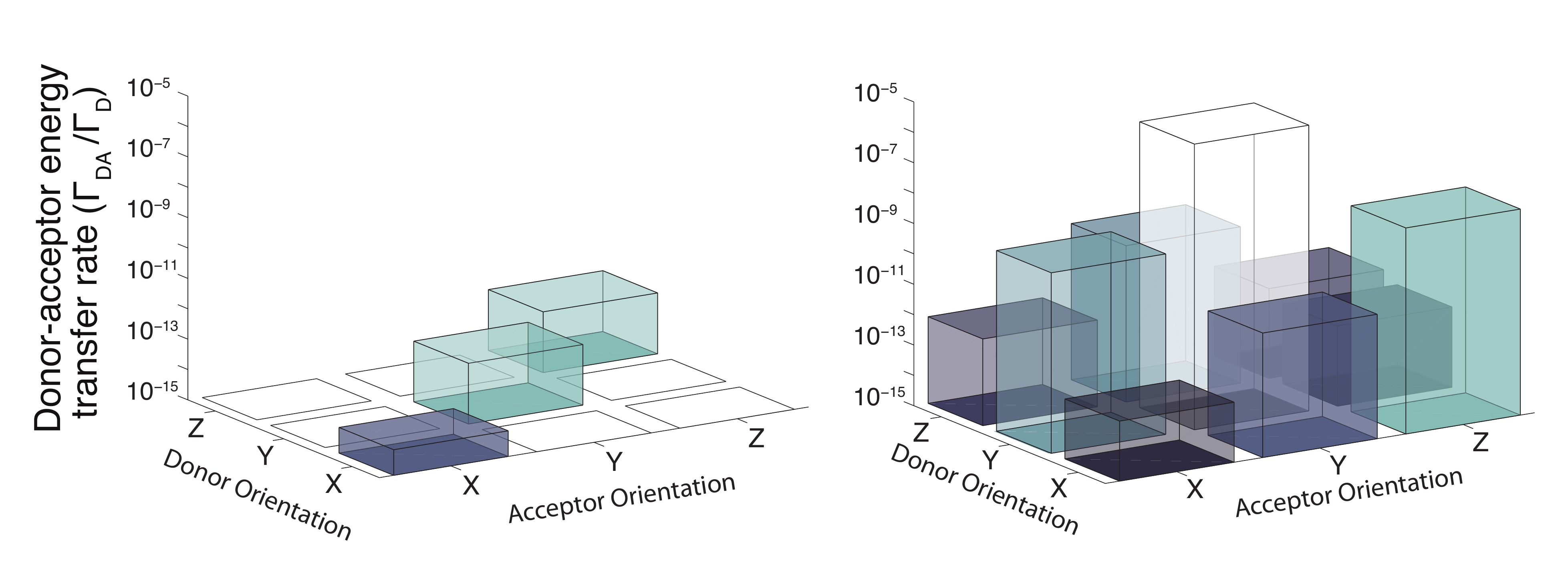} 
\caption{\label{Fig.3} (Color online) Energy transfer tensor for different donor-acceptor mutual orientations, at 1 $\mu$m distance at a wavelength of 800 nm. Vacuum (left) shows only diagonal terms, as expected from the far-field mediated interaction between donor and acceptor. Alternatively, for the hybrid environment-mediated energy transfer (right), off-diagonal elements are present which lead to a more efficient energy transfer. Units of the vertical axis are transfer events per donor emission event. }
\end{center}
\end{figure}

In order to highlight the energy transfer spatial dependence, we define a new figure of merit $\xi(\mathbf{r}_D,\mathbf{r}_A)$, which represents the enhancement of the normalised-to-donor-emission energy transfer rate in the presence of inhomogeneous environments compared to vacuum:

\begin{equation}
\xi(\mathbf{r}_D,\mathbf{r}_A)  \equiv \frac{f^{inh}}{f^{vac}} =  \frac{\Gamma^{inh}_{DA}/\Gamma^{inh}_{D}}{\Gamma^{vac}_{DA}/\Gamma^{vac}_{D}}
					 = \frac{\Gamma^{inh}_{DA}}{\Gamma^{vac}_{DA}} \frac{1}{P_{D}},
\end{equation}
where the term $P_{D} = \Gamma_D^{inh}/\Gamma_D^{vac}$, accounts for the Purcell enhancement at the donor location. $\xi(\mathbf{r}_D,\mathbf{r}_A)$, when evaluated for a homogeneous environment with constant refractive index $n>1$, is equal to \emph{$P_D$} at all distances R.  %

For the waveguide geometry shown in Fig. 2 we obtain Purcell factors $P_x \sim 2$, $P_y  \sim 2$ and $P_z  \sim 3$, for dipoles oriented along the three cartesian axis, purely related to the presence of the silicon nitride-air interface. The waveguide is mildly dispersive, yielding Purcell enhancement values that do not change considerably as a function of the wavelength. On the contrary, the radiative enhancement of the donor emission rate around the optical antenna is remarkably high and dispersive \cite{antennas} (see Supplementary Information): an emitter with transversal orientation along the antenna experiences a considerable emission rate modification ($P_y \sim 330$), since it couples efficiently to the near-field dipolar mode of the antenna. Transversally oriented emitters couple to the transversal modes of the antenna, which are not efficiently excited when the dipole is located at the end of the rod antenna, making their associated Purcell factors smaller than in the longitudinal case ($P_x  \sim 6$, $P_z  \sim 7$). Despite this modest Purcell enhancements, we calculated the transfer rate (at a distance of 1 $\mu$m) due to the \emph{off-diagonal} Green's function contributions between a x-directed donor and a y-directed acceptor to be almost 4-orders of magnitude higher than between two dipoles oriented along the same x-axis (Fig. 3). Long-range energy transfer, mediated by transversal photons, is instead forbidden for orthogonally-oriented dipoles in vacuum, since their far-field radiation patterns have orthogonal polarisation.

\begin{figure}
\begin{center}
\includegraphics[width=6.0in]{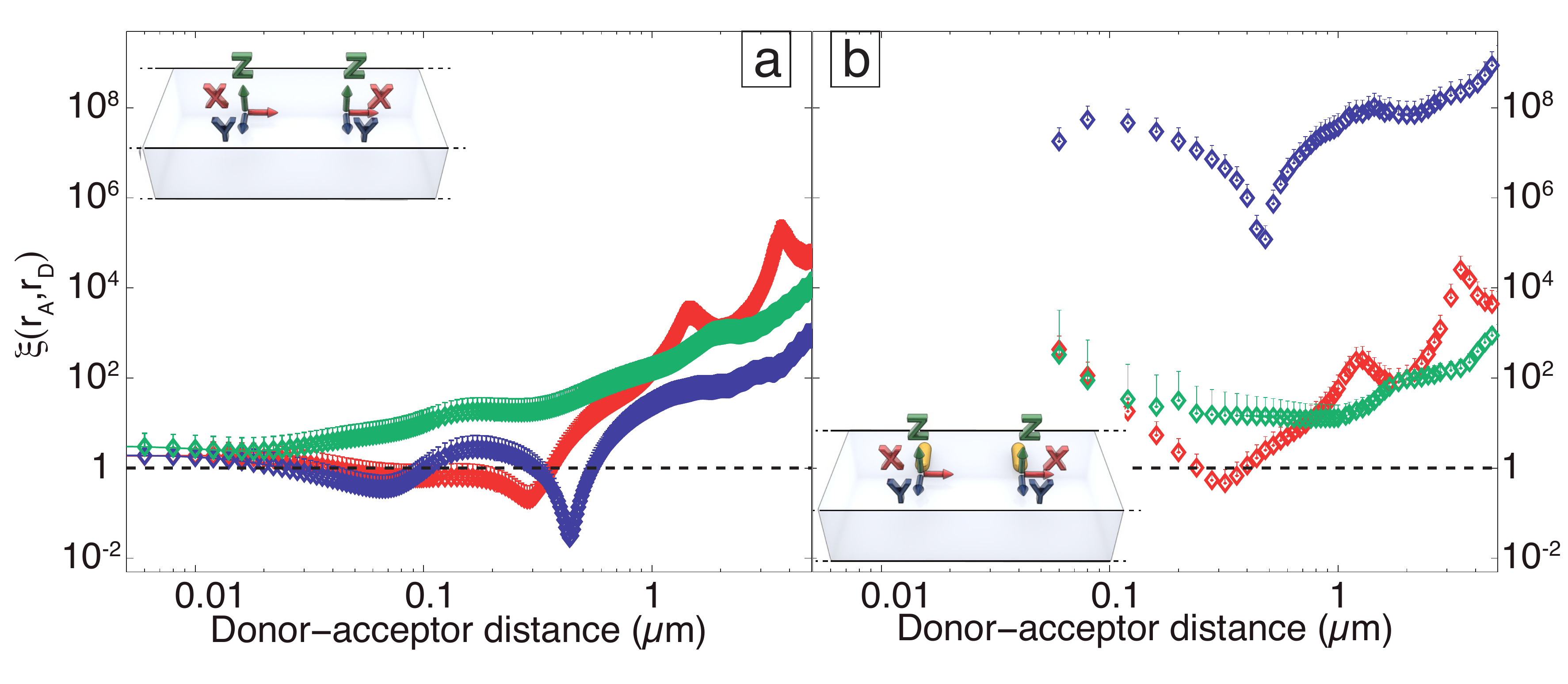} 
\caption{\label{Fig.4} (Color online) Distance dependence of $\xi(\mathbf{r}_D,\mathbf{r}_A)$ for dipole orientations along $X_D-X_A$ (orange), $Y_D-Y_A$ (violet) and $Z_D-Z_A$ (green). The graph plots $\xi(r_D,r_A)$ for the case of energy transfer mediated by a dielectric waveguide (a) and by optical antennas on a dielectric waveguide (b). The efficiency of energy transfer is maximised for y-oriented dipoles in a hybrid antenna-waveguide geometry. After around 10 nm $\xi$ starts deviating from its initial value. }
\end{center}
\end{figure}

In Fig. 4 we study the distance dependence of $\xi(\mathbf{r}_D,\mathbf{r}_A)$: for the waveguide-only case and at short donor-acceptor distances below $\sim 10$ nm $\xi(\mathbf{r}_D,\mathbf{r}_A)$ is close to unity (dotted line), within our numerical accuracy, for all different dipole orientations, with only an offset related to their different Purcell factors. We checked that when normalised by the Purcell factor the three curves are identically equal to unity for very short distances. Deviations are instead visible for  donor-acceptor separations larger than $\sim 10$ nm. Within our numerical uncertainty, the values of $\xi$ at short distances for the waveguide alone (Fig. 4a) point towards the existence of a correlation between the energy transfer rates and the LDOS, even if further work is required to investigate such short distances and the eventual functional LDOS dependence.
Instead, as already pointed out by Blum et al. \cite{Blum}, energy transfer would not depend on the LDOS if one can assume that $\mathrm{Im}[\mathbf{G}(\mathbf{r}_D,\mathbf{r}_A, \omega)]$ can be approximated by the zero-order Taylor term $\mathrm{Im}[\mathbf{G}(\mathbf{r}_D,\mathbf{r}_D, \omega)]$, with both donor and acceptor experiencing the same Purcell effect. 
 For inhomogeneous environments with LDOS values rapidly changing over the donor-acceptor distances, further terms of this Taylor expansion would be needed to accurately describe the problem, leading to terms depending on spatial derivatives (gradients) of the LDOS. These higher order terms become more important with increasing distance between donor and acceptor. In our particular case of a waveguide mediated energy transfer, deviations from the initial values of $\xi$ start occurring for donor-acceptor distances of $\sim 10$ nm. Deviation for similar distances have been found in recent experimental studies of plasmonic nano-apertures \cite{J.Wenger}. As shown in Fig. 4,  $\xi(\mathbf{r}_D,\mathbf{r}_A)$ presents an even larger increase for larger distances ($> 100$ nm). The waveguide in our calculations sustains a propagating mode, therefore the enhancement parameter $\xi(\mathbf{r}_D,\mathbf{r}_A)$ is expected to increase quadratically with distance as the energy transfer saturates, because the transfer rate in vacuum decreases as $R^{-2}$. The modulations of $\xi(\mathbf{r}_D,\mathbf{r}_A)$ are due to interference of the light coupled into the waveguide, similar to what was described in Fig. 2a. 

Fig. 4b plots $\xi(\mathbf{r}_D,\mathbf{r}_A)$ for the hybrid photonic structure. In this case, the energy transfer enhancement at a donor-acceptor distance of $1$ $\mu$m is increased by 6 orders of magnitude for dipoles orientated along the long axis of the antennas (blue diamonds) when compared to the waveguide-only case. For dipole orientations transversal to the long antenna axis, there is a $\sim 10$-fold decrease of the energy transfer at the same distance (1 $\mu$m), as compared to the waveguide-only case. This decrease is due to the near-field mismatch between the longitudinal dipolar field of the optical antenna and that of the emitter, which is oriented perpendicular to the antenna; this mismatch can be easily compensated by changing the position of the dipoles. 

\section{Conclusion}

To conclude, we have presented enhanced long-range energy transfer mediated by a complex hybrid photonic environment where donor and acceptor are coupled to metallic antennas linked by a dielectric waveguide. We report $2\cdot10^{-6}$ absolute energy transfer to donor decay rate, which suggests that experimental verification could be done for donor-acceptor distances of several wavelengths. These results are obtained using a newly developed numerical technique based on the FDTD method, that can be applied to arbitrary geometries. Moreover, our results points to a complex dependence of energy transfer on the photonic environment, beyond the simple donor Purcell enhancement, which increases for larger donor-acceptor distances. The design we propose, of hybrid dielectric and plasmonic components, could be further improved by using more complex antennas to better couple the emitters and the waveguide mode and could open new avenues for single-photon exchange between emitters for future quantum technologies.

\section{Acknowledgments}
We thank N. de Sousa and R. Carminati for stimulating discussions.  This research was funded by the Engineering and Physical Sciences Research Council (EPSRC), the European Commission (ERC Advanced Grant 247330-Nano Antennas), a Leverhulme Trust Research Grant and Fundaci\'o CELLEX (Barcelona). PMdR acknowledges financial support from Spanish Government MINECO-FPI grant and European Science Foundation under the PLASMON-BIONANOSENSE Exchange Grant program.

\section{References}


\end{document}